\documentclass[amssymb, twocolumn, amsmath, showpacs, prl,superscriptaddress, notitlepage]{revtex4-1}
\usepackage{}

\usepackage{graphicx,graphics,epsfig,subfigure,times,bm,bbm,amssymb,amsmath,amsfonts,amsthm,mathrsfs,MnSymbol}
\usepackage[matrix,frame,arrow]{xypic}
\usepackage[pdfstartview=FitH]{hyperref}
\usepackage[pdftex]{color}
\usepackage{hyperref}
\usepackage{enumerate}
\usepackage{epstopdf}
\usepackage{pdfsync}
\usepackage{relsize}

\newtheorem{definitionenv}{Definition}
\newtheorem{remarkenv}[definitionenv]{Remark}

\newtheorem{mytheorem}{Theorem}

\newcommand{\bes} {\begin{subequations}}
\newcommand{\ees} {\end{subequations}}
\newcommand{\bea} {\begin{eqnarray}}
\newcommand{\eea} {\end{eqnarray}}

\newcommand{\beq}{\begin{equation}}

\newcommand{\wt}[1]{\mathrm{wt}(#1)}
\newcommand{\eeq}{\end{equation}}

\newcommand{\ignore}[1]{}
\newcommand{\mc}[1]{\mathcal{#1}}

\def\>{\rangle}
\def\<{\langle}

\newcommand{\ket}[1]{|#1\rangle}

\newcommand{\cC}{\mathcal{C}}

\begin{document}
\title{Teleportation-based Fault-tolerant Quantum Computation in Multi-qubit Large Block Codes}

\author{Todd A. Brun}
\email{tbrun@usc.edu}
\author{Yi-Cong Zheng}

\email{
yicongzh@usc.edu}
\affiliation{Department of Electrical Engineering,  University of Southern California, Los Angeles, California, 90089\\}
\affiliation{ Center for Quantum Information Science \& Technology, University of Southern California, Los Angeles, California, 90089\\}

\author{Kung-Chuan Hsu}
\affiliation{Department of Electrical Engineering, University of Southern California, Los Angeles, California, 90089\\}
\affiliation{ Center for Quantum Information Science \& Technology, University of Southern California, Los Angeles, California, 90089\\}

\author{Joshua Job}
\affiliation{Department of Physics and Astronomy, University of Southern California, Los Angeles, California, 90089}

\affiliation{ Center for Quantum Information Science \& Technology, University of Southern California, Los Angeles, California, 90089\\}
\author{Ching-Yi Lai}
\affiliation{Centre for Quantum Computation and Intelligent Systems (QCIS), University of Technology, Sydney, NSW 2007, Australia}

\begin{abstract}
A major goal for fault-tolerant quantum computation (FTQC)  is to reduce the overhead needed for error correction.
One approach is to use block codes that encode multiple qubits, which can achieve significantly higher rates for the same code distance than single-qubit code blocks or topological codes.
We present a scheme for universal quantum computation using multi-qubit Calderbank-Shor-Steane (CSS) block codes, where codes admitting different transversal gates are used to achieve universality, and logical teleportation is used to move qubits between code blocks.
All circuits for both computation and error correction are transversal. We also argue that single shot fault-tolerant error correction can be done in Steane syndrome extraction. Then, we present estimates of information lifetime for a few possible codes, which suggests that highly nontrivial quantum computations can be achieved at reasonable error rates, using codes that require significantly less than 100 physical qubits per logical qubit.

\end{abstract}

\maketitle


Quantum computers (QCs) are extremely vulnerable to errors during the computation process.
Theory has shown that if errors of each type are sufficiently local, and their rates are small enough to fall below a threshold, it is possible to carry out quantum computations of arbitrary size with arbitrarily small error, by so-called \emph{fault-tolerant} methods using quantum error-correcting codes (QECCs)~\cite{DivencenzoFTPhysRevLett.77.3260,Gaitan:2008:CRC,QECbook:2013}. Since the threshold theorem has been established, a number of fault-tolerant schemes have been proposed, including but not limited to those introduced
in Refs.~\cite{Gottesman:9705052,Gottesman:1998:127,KnillFTNature, Raussendorf:2007:190504,Raussendorf:2007:199,Yi-Cong_PhysRevA.89.032317}.

From the practical viewpoint, we say that an FTQC scheme is \emph{fault-tolerant} if the logical error rates for each elementary logical operation in the scheme are sufficiently low so that a quantum algorithm can be executed with a high probability of success.
Typically,  a practical quantum  algorithm using $K$ qubits and $Q$ elementary steps has $KQ\gg 10^{10}$, so the logical error rate for each logical operation should be much less than $10^{-10}$ \cite{van2013blueprint}.
However, most FTQC schemes require enormous  overhead to achieve this rate by increasing either the \emph{concatenation levels} for concatenated codes or the \emph{distance} (hence the size) for topological codes.
As a result, a logical qubit is encoded in thousands of physical qubits~\cite{KnillFTNature,Folwer2012PhysRevA.86.032324,van2013blueprint,ching-yiKnill}.

\begin{figure}[!ht]
\centering\includegraphics[height = 45mm,width=55mm]{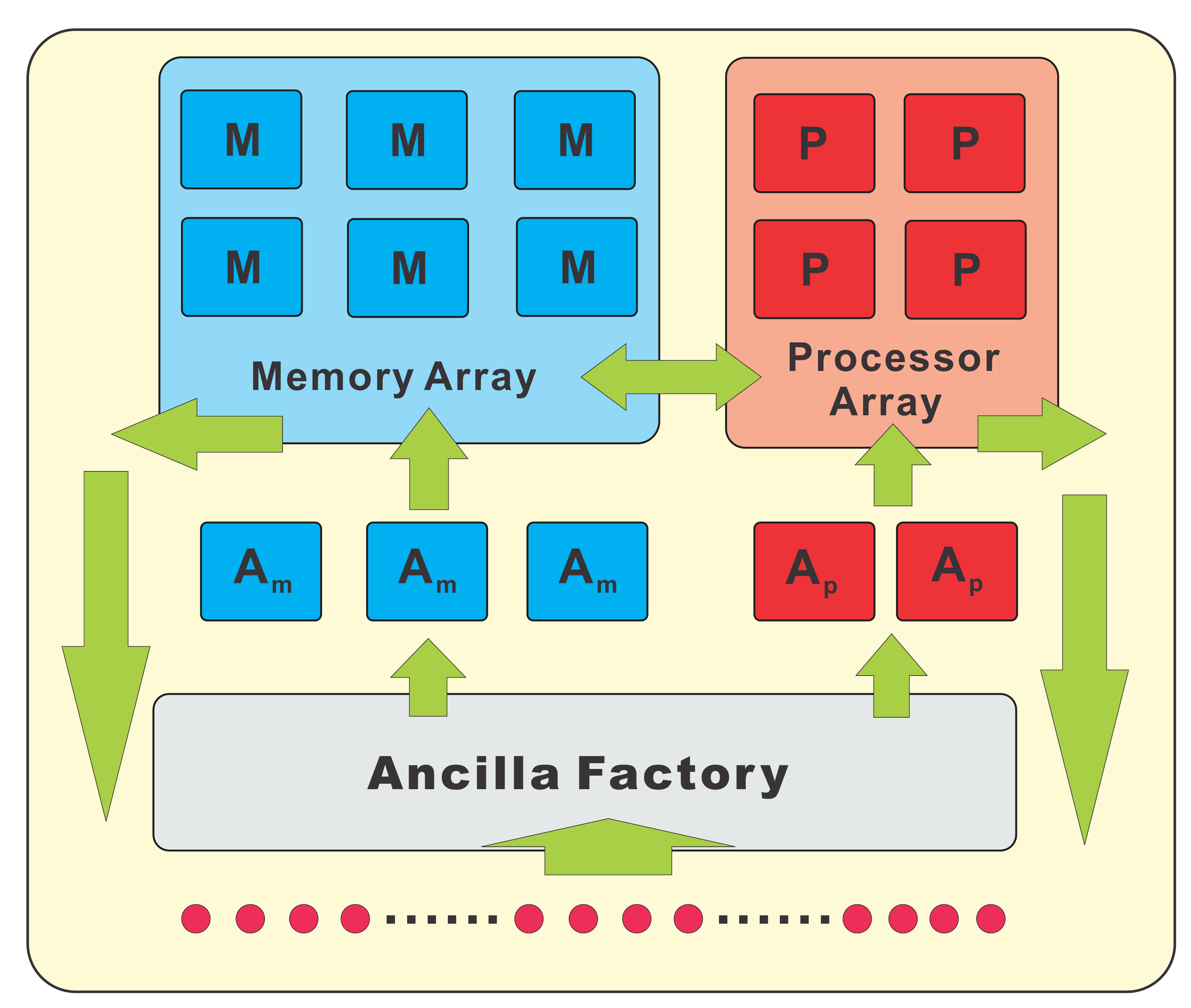}
\caption{\label{fig:diagram}The architecture of our teleporation-based FTQC scheme. }
\end{figure}

It was observed by Steane more than a decade ago that multi-qubit block codes can achieve significantly higher  code rates for comparable error protection ability,
but  logical gates in these codes  are quite difficult to implement ~\cite{steane1999efficient_Nature,Steane:2003:042322}.
In this paper, we propose a scheme that exploits the advantages of  multi-qubit block codes.
Similar to von Neumann architecture, our scheme has three components
as shown in Fig.~\ref{fig:diagram}: a \emph{memory} array of  $[[n,k,d]]$ CSS code blocks $(\cC_{m})$~\cite{Calderbank:1996:1098, Steane:1996:793} with $k>>1$;
a \emph{processor} array of an $[[n',1,d']]$ quantum code blocks $(\cC_p)$ that support a transversal  $T$ $(\pi/8)$  gate  (or other non-Clifford gate); and an \emph{ancilla factory}
that continuously produces a variety of fresh logical ancillas for error correction, teleportation, and {logical operator measurement}.
Another feature of our scheme  is that  magic state distillation~\cite{Folwer2012PhysRevA.86.032324},
which usually dominates the overhead of an FTQC scheme,
is not required as in Refs.~\cite{paetznick2013universal,jochym2014using,ADP14}.

\emph{Universal FTQC}.--
Quantum information is stored in the memory array,
and error corrections using Steane syndrome extraction~\cite{steane1997active} are constantly performed.
Logical Clifford operations can be implemented by measuring sequences of logical operators on the $\cC_m$ code blocks in the memory array.
We will show that measuring  logical Pauli operators of $\cC_m$ can be combined with error correction if some particular ancilla states are available.
To implement a logical $T$ gate on a particular logical qubit,
that logical qubit will be teleported to a $\cC_p$ code block, where a transversal $T$ gate is performed,
and  then  teleported back to its original memory block.
Error corrections between these operations  can be performed if necessary.
Again, provided with some  particular ancilla states, it is possible to simultaneously measure the logical operators and implement logical teleportation between  $\cC_m$ and $\cC_p$ code blocks.
Thus, universal quantum computation is achieved.
Also, this scheme needs only ancilla preparation, transversal circuits, and single-qubit measurements, and thus it is intrinsically fault-tolerant.
Details of these logical operations will be given below.

It is evident that a large number of clean ancillas of various types are required in this scheme.
Fortunately, these logical ancillas are stabilizer states. They can be prepared by using  quantum circuits of Clifford gates only and then distilled~\cite{Ching-yiAncillaDistillation}.
The distillation procedure is more like entanglement distillation~\cite{Bennett:1996:722} than magic state distillation, and has some advantages.
For magic state distillation, there is a probability of failure, where everything has to be discarded, and it may need several iterations;
while stabilizer states can be measured, and logical errors, if detected, can be corrected.
We will not go through the details of this logical ancilla distillation in this paper,
but simply assume that we have an ancilla factory capable of preparing all the ancillas with high fidelity.

\begin{figure}[!ht]
\centering\includegraphics[width=85mm]{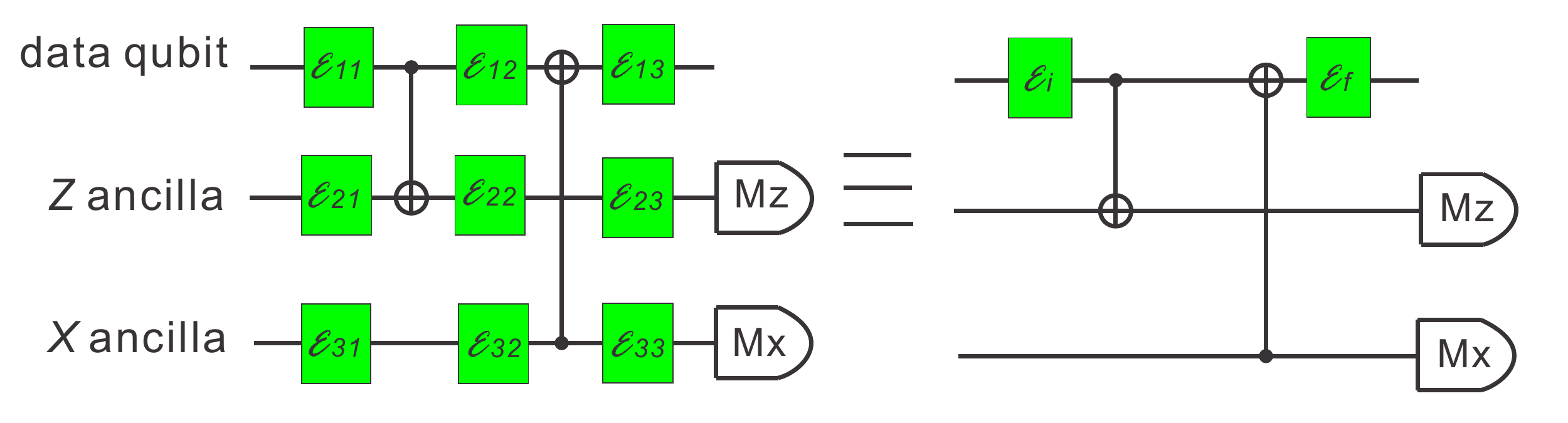}
\caption{\label{fig:effective_error} The circuit for Steane syndrome extraction and its effective error model.}
\end{figure}

\emph{Steane Syndrome Extraction}.--
First let us briefly review the Steane syndrome extraction, which leads to the other operations.
It is used to measure the stabilizer generators for the $[[n,k,d]]$ CSS code $\cC_m$ (similarly for $\cC_p$ if necessary) in our scheme.
The procedure is as follows:
1) Prepare two ancilla states in the same code $\cC_m$ (or $\cC_p$) where all logical qubits are set to the states $\ket{0}_L^{\otimes k}$
and $|+\>_L^{\otimes k}$
(called the $X$ and $Z$ ancillas respectively),
for $Z$ and $X$ error syndrome measurements, respectively.
2) Perform a transversal CNOT from the information block to the $Z$ ancilla.
3) Perform a transversal CNOT from the $X$ ancilla to the information block.
4) Do single-qubit measurements on the $X$ and $Z$ ancilla qubits in the $X$ and $Z$ basis, respectively.
Collecting the measurement outcomes and multiplying the correct subset of $\pm1$ results together reveals the eigenvalue of each stabilizer generator, and hence the error syndrome.
This procedure is shown as the left circuit in Fig. \ref{fig:effective_error}.

\emph{Measuring logical operators}.--
Steane has shown that measuring a logical $\bar{X}_{\bf u}$ or $\bar{Z}_{\bf u}$ can be combined with the  recovery operation in Steane syndrome extraction~\cite{steane1999efficient_Nature},
where $\bf{u}$ is a binary $k$-tuple indicating which logical qubits are operated.
If logical $\bar{X}_{\bf u}$ is to be measured, the prepared $X$ ancilla $\ket{{\bf 0}}_L$ at step 1) is replaced with $\ket{{\bf 0}}_L+\ket{{\bf u}}_L$, which is also a stabilizer state,
and the rest of the steps are the same as before.
As an example, suppose we wish to measure $\bar{X}_i\bar{X}_j$. Then the logical qubits $i$ and $j$ of the $X$ ancilla are prepared 
 in the state  $\ket{\Phi_+}_L= \frac{1}{\sqrt{2}}(\ket{0_i0_j}_L+\ket{1_i1_j}_L)$,   
which is a joint $+1$ eigenstate of $\bar{X}_i\bar{X}_j$ and $\bar{Z}_i\bar{Z}_j$,
and the other logical qubits are prepared in the state $\ket{0}_L$.
This logical operator measurement is protected by the classical error-correcting code from which $\cC_m$ (or $\cC_p$) is built up.

Next we generalize this method to products of logical $\bar{X}$  and $\bar{Z}$ operators.
To illustrate, we show how to measure logical operators of the form $\bar{X}_i\bar{Z}_j$ on logical qubits $i$ and $j$.
This will allow us to do any Clifford gate as we shall see.
In this case, if $i\neq j$,
the $X$ and $Z$ ancillas at step 1) are prepared in a particular entangled logical state:
logical qubit $i$ of the $X$ ancilla and logical qubit $j$ of the $Z$ ancilla are prepared in the entangled state $|\Omega_{ij}\>_L = 1/2\left(|0_i0_j\>_L + |0_i1_j\>_L + |1_i0_j\>_L - |1_i1_j\>_L\right)$,
which is the joint $+1$ eigenstate of  $\bar{X}_i\bar{Z}_j$ and $\bar{Z}_i\bar{X}_j$,
while the other logical qubits of the $X$ or $Z$ ancillas are prepared in the state $|0\>_L$ or $|+\>_L$, respectively. If $i=j$, we need to prepare the ancilla as a joint $+1$ eigenstate of $\bar{Y}_i\bar{Z}_i$ and $\bar{Z_i} \bar{X}_i$.
Again, these joint $2n$-qubit states are a stabilizer states.
It is similarly possible to measure any logical operator $\bar{X}_{\bf u}\bar{Z}_{\bf v}$, by preparing more complicated ancilla states.
It is important to emphasize that these  measurements are combined with error correction, as in the original Steane syndrome extraction.

\emph{Logical teleportation}.--
We show that logical qubits can be teleported between arbitrary code blocks (of $\cC_m$ or $\cC_p$).
To perform a non-Clifford quantum gate on a logical qubit (or qubits) of a $\cC_m$ memory block, the target logical qubit is teleported to a $\cC_p$ processor block that allows the non-Clifford gate to be implemented transversally.
One can think of the two code blocks as a part of a larger code. Suppose the logical qubits of $\cC_m$ have associated with them pairs of logical operators ($\bar{X}_1$, $\bar{Z}_1$), ($\bar{X}_2$, $\bar{Z}_2$),..., ($\bar{X}_k$, $\bar{Z}_k$), which are all Pauli operators.
We reserve logical qubit 1 as a buffer qubit used in teleporting qubits. Suppose the logical operators of the $ [[n', 1,d']]$ code $\cC_p$  are labeled as ($\bar{X}_0$, $\bar{Z}_0$).
Here is the procedure to teleport logical qubit $j$ from the storage block to the processing block:
1) Measure the operators $\bar{X}_0\bar{X}_1$ and $\bar{Z}_0\bar{Z}_1$. This prepares a logical Bell state between the processor block and the buffer qubit.
2) Measure the operators $\bar{X}_1\bar{X}_j$ and $\bar{Z}_1\bar{Z}_j$. This does a logical Bell measurement on the buffer qubit and qubit $j$, and teleports qubit $j$ to the processor block.
3) If necessary, apply a logical Pauli operator to the processor  block to correct the state.
One would generally need to do correction before applying the non-Clifford gate.
4) Apply the non-Clifford gate by a transversal circuit on the processor block.
5) Measure the operators $\bar{X}_0\bar{X}_1$ and $\bar{Z}_0\bar{Z}_1$. This prepares a logical Bell measurement on the processor
block and the buffer qubit, and teleports the transformed qubit back to logical qubit $j$ of the memory block.
6) If desired, apply a logical Pauli operator to the memory block to correct.
\begin{figure}[!htp]
\centering\includegraphics[width=80mm]{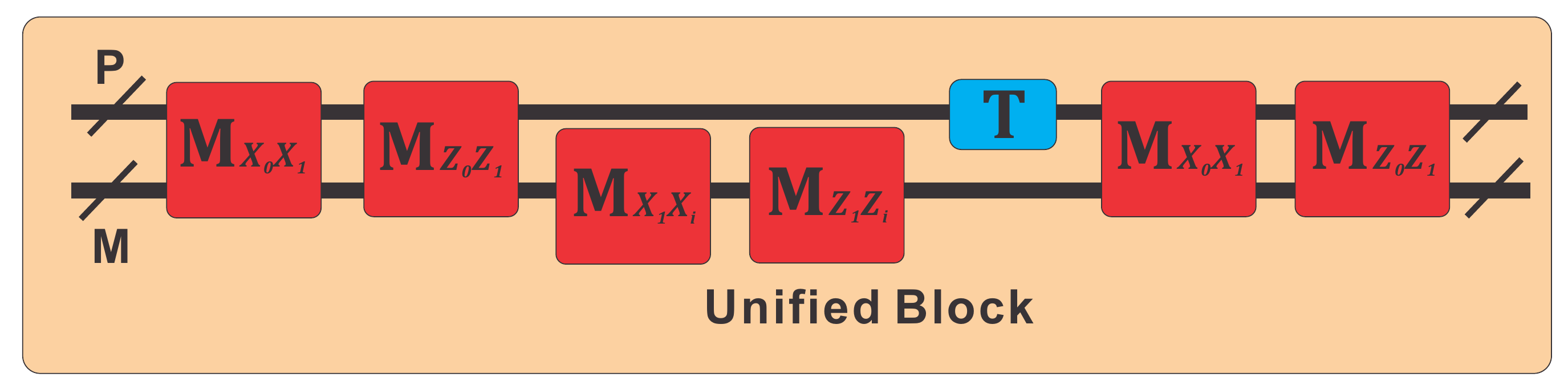}
\caption{\label{fig:logical_teleportation} (Color online) Diagram of the logical $T$ gate using logical state teleportation. The red blocks represent joint measurements of logical qubits, and the blue one represents bitwise $T$ or $T^\dag$ applied to the processor block.}
\end{figure}
The procedure is illustrated in Fig.~\ref{fig:logical_teleportation} for a logical $T$ gate.
The steps of measuring logical operators are similar to what was described previously, except that
the $X$ and $Z$  ancillas are prepared in a logical entangled state between $\cC_m$ and $\cC_p$.
Logical teleportation can also be applied, of course, to move logical qubits between memory blocks.

\emph{Logical Clifford gates.}--
Clifford gates can be performed within a memory block solely by
measuring logical operators, which is like a kind of simplified logical teleportation~\cite{Gottesmanpriviate}.
We demonstrate how to perform a logical Hadamard gate and a logical CNOT gate, while the Phase gate and the SWAP gate are left to
the supplementary material.
Suppose we wish to perform a logical Hadamard gate on logical qubit $i$ of a $\cC_m$ block in the state $|\psi\>_L$. Logical qubit 1 is reserved as a buffer qubit in the state $|0\>_L$.
We do the following two measurements:
1) measure $\bar{X}_1\bar{Z}_i$;
2) measure $\bar{X}_i$.
Logical qubit 1 will be left in the state $\bar{H}|\psi\>_L$,
up to a Pauli correction $\bar{X}_1$, $\bar{Y}_1$ or $\bar{Z}_1$ on logical qubit 1, according to the measurement outcomes.
Logical qubit $i$ will be left in the state $|+\>_L$ or $|-\>_L$
and can be reset to $|0\>$ as a new buffer qubit, if desired, by measuring $\bar{Z}_i$ and applying an $\bar{X}_i$ correction if necessary.
A CNOT can be performed similarly.
Suppose we wish to perform a CNOT from logical qubits $i$ to $j$ of a $\cC_m$ block,
and again, suppose logical qubit 1 is a buffer qubit in state $|0\>_L$.
Here is the procedure: 1) Measure $\bar{X}_1\bar{X}_j$; 2) Measure $\bar{Z}_i\bar{Z}_j$; 3) Measure $\bar{X}_i$.
This does a CNOT from logical qubits $i$ to $j$ (up to a Pauli correction), shifts them to logical qubits $j$ and 1, respectively, and moves the buffer qubit to qubit $i$ in the state $|\pm\>_L$.
We can build any Clifford unitaries from Hadamard, Phase, CNOT, and SWAPs.
But a complicated Clifford unitary can also be done directly by measuring more
complicated combinations of logical operators. Each of these measurements requires the preparation of a particular ancilla state.
Therefore a tradeoff exists between the efficiency of enabling a larger set of possible Clifford operations
and the complexity of having to prepare and distribute more kinds of ancillas.

Note also that it is not necessary to apply the Pauli corrections; we can just keep track of them and how they are transformed by Clifford unitaries (the `Pauli frame'). But if we do wish to correct, that is a transversal operation as well.

\emph{Error model.}--
Steane syndrome extraction and its variations are used throughout the scheme.
There are at least four kinds of errors in this scheme: memory errors in the code blocks, physical gate errors, faulty ancilla preparation, and measurement errors.
We model errors in physical gates, ancilla preparations, and measurements by treating them as perfect operations followed or preceded by Pauli errors.
In this paper  we represent the physical noise model as depolarizing errors.
At each time step, every physical qubit \emph{independently} undergoes a Pauli error $X$, $Y$ or $Z$ with probability $\epsilon/3$
or remains unchanged with probability $1-\epsilon$,  where $\epsilon$ is called the memory error rate.
We treat ancilla preparation as perfect followed by each qubit of an ancilla block {independently} suffering the same depolarizing errors with rate $r$ afterwards.
Similarly, we treat each single-qubit gate as perfect, followed by a depolarizing error with rate $p_{g_1}$ for single-qubit gates.
For a two-qubit gate, it is modeled as a perfect gate followed by
one of the 15 possible single- or two-qubit error from  $IX$, $IY$, $IZ$, $XI$, $XX$, $XY$, $XZ$, $YI$, $YX$, $YY$, $YZ$, $ZI$, $ZX$, $ZY$, and $ZZ$
with equal probability $p_{g_2}/15$,
or no error with probability $1-p_{g_2}$.
Finally, the measurement of a single physical qubit has a classical bit-flip error with  probability $p_m$ (or equivalently, an $X$ or $Z$ error preceding a measurement in the $Z$ or $X$ basis, respectively). Note that we do not expect the \emph{form} of the errors to greatly affect the performance; but the assumption of independence across qubits is very important.

The measurement outcomes in Steane syndrome extraction can be erroneous  due to either imperfect measurements or errors during the circuit. This makes error analysis difficult.
Traditionally this is handled by repeated syndrome measurements.
Herein we show that syndrome measurement can be done in a \emph{single shot}. Actually, all errors during the syndrome extraction process can be mapped to errors occurring on the data qubits
before and after the process, so that the ancillas, gates and qubit measurements in the circuit can be regarded as error-free,
as illustrated in Fig.~\ref{fig:effective_error}.
This is formally stated as the following theorem.
\begin{mytheorem}\label{lemma:effective_error}
(Effective error) During the process of imperfect Steane syndrome extraction and its variations, if errors in the same block (memory, processor, or ancilla) are uncorrelated,
then the errors are equivalent to effective errors acting
only on the \emph{data} qubits before and after the process. 
\end{mytheorem}
\noindent  The idea is to commute errors forward or backward in the circuit. This theorem is applicable to quite general independent noise models and not just the depolarization channel we focus on in this paper.
Consequently we only have to decode the effective error on the data qubits at each syndrome measurement. Since syndromes are measured at every step of the process, this avoids repeated syndrome measurement and greatly reduces the time overhead for syndrome extraction and implementing logical gates. On the other hand, it can also eliminate potential errors caused by repeated measurements.
The estimated effective error will be a Pauli operator and can
either be corrected instantly,
or kept track of in the cumulative Pauli frame on the data qubits.   
After every round, there will generally be a residual error that has not yet been detected, and that differs from the current error estimate.
However, this is not a problem so long as the weight of this residual error is always small compared to the distance of the code.
At the same time, it greatly simplifies the analysis of error propagation.

For example, if we assume $\epsilon=p_m=p_{g_1}=p_{g_2}\triangleq p$, the effective model on the data block at each time step can be approximated by
\beq
\mathcal{E}_{\text{tot}}[\rho]\approx  (1-11p)\rho+\frac{71}{15}pX\rho X+\frac{71}{15}pZ\rho Z + \frac{23}{15}pY\rho Y.
\eeq
(Details of this approximation and the proof of Theorem~\ref{lemma:effective_error} are given in the supplementary materials.)
For simplicity, we choose the error model to be a depolarizing error with  rate $p_{\text{eff}}=\left(71/5\right) p$ in the following simulation, where $p$ is the underlying physical error rate.



\begin{figure}[!ht]
\centering\includegraphics[width=80mm]{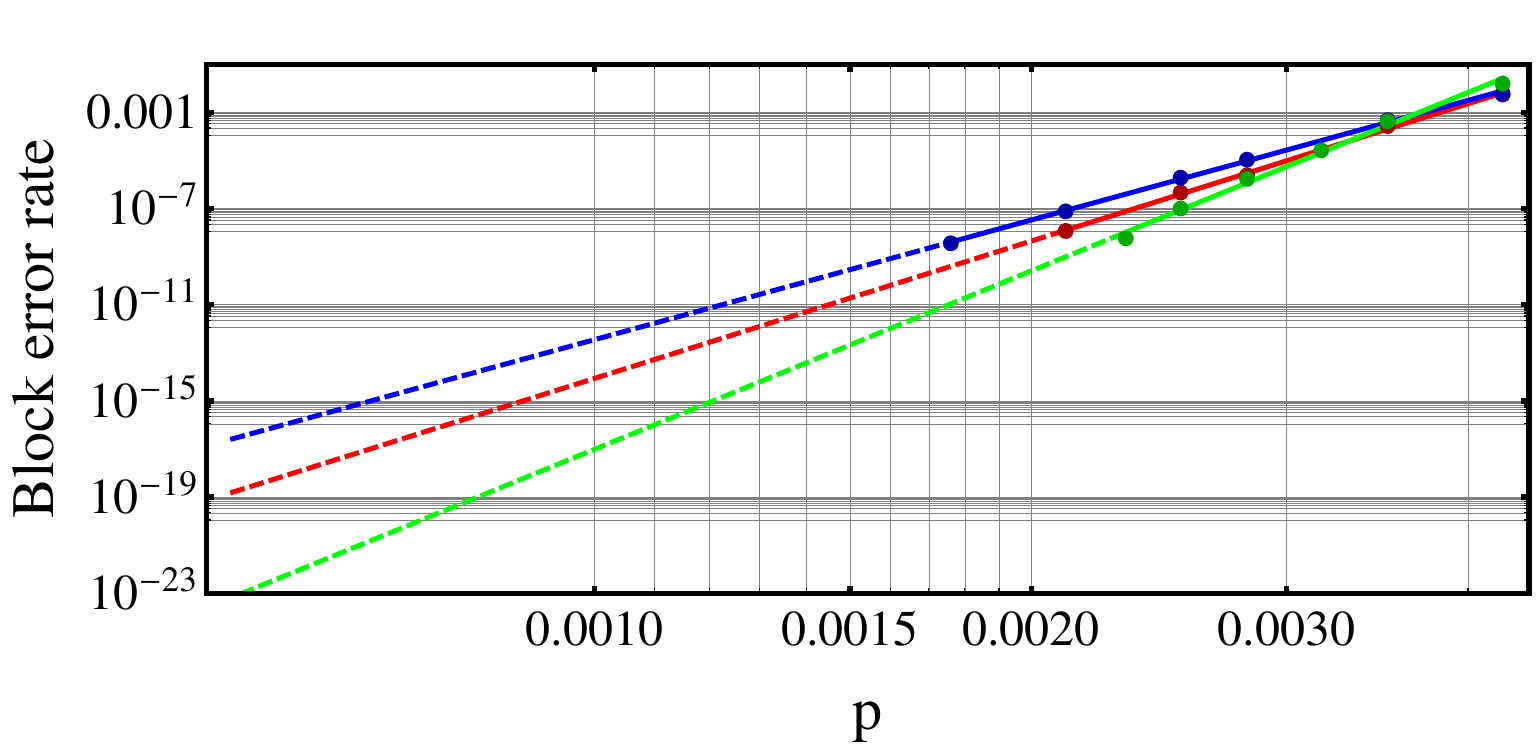}
\caption{\label{fig:memory_block} (Color online) The logical error rate of the memory blocks for the [[2047,23,77]] code (blue), [[2921,57,77]] code (red) and [[5865,143,105]] code (green) versus physical error rate $p$.
The number of samples for each point is up to $4\times 10^8$. The dashed lines are from extrapolation of linear fitting.  }
\end{figure}

\emph{Estimate of the logical error rate.}--
The error rate for each logical step is determined by the failure rate of decoding for the effective error process.
The performance and the physical resources of our scheme will depend heavily on the choice of quantum codes for the memory and processor blocks.
For memory blocks, desirable properties include: 1) high distance,
2) good code rate, and 3) an efficient decoding algorithm. As preliminary research, we study three large block codes obtained by concatenating a medium-sized block code with a high-distance single-qubit code.
By concatenating the $[[89,23,9]]$, $[[127,57,11]]$, and $[[255,143,15]]$ quantum BCH codes~\cite{Grassl:1999:207}
with the $[[23,1,7]]$ quantum Golay code, we obtain CSS codes with parameters $[[2047,23,63]]$, $[[2921,57,77]]$ and $[[5865,143,105]]$,  respectively.
All three block codes on average encode a single logical qubit in less than $100$ physical qubits.
The Golay code is decoded using the Kasami error-trapping decoder~\cite{kasami1964decoding},
and the BCH codes are decoded using the Berlekamp-Massey algorithm~\cite{berlekamp1968algebraicBCH, massey1969BCH}.
Fig.~\ref{fig:memory_block} shows the simulation of logical error rate for a memory block using Monte-Carlo simulation.
However, the simulation complexity is too high to go beyond $p_{\text{eff}}\lesssim 10^{-2}$,
even with the Titan supercomputing resource.
Thus, we use linear extrapolation to estimate that region, and find that at $p_{\text{eff}}=0.007$ (corresponding to $p=5\times10^{-4}$)
the logical error rates are less than $10^{-15}$  for all the three codes.

We can also derive  an upper bound on the expected logical error rate of an  $[[n,k,d]]$ code.
Since all errors of weight up to $t =\lfloor \frac{d-1}{2}\rfloor$ can be corrected, we pessimistically assume that any error of higher weight would lead to a logical error.
The approximate logical error probability is $P_n(p)=\sum_{w=t+1}^{n}\binom{n}{w}p^w (1-p)^{n-w}$. At effective error rate $p_{\text{eff}}=0.007$, we get $P_{89}(P_{23}(p_{\text{eff}}))=1\times 10^{-16}$, $P_{127}(P_{23}(p_{\text{eff}}))
=2.5\times 10^{-19}$, and $P_{255}(P_{23}(p_{\text{eff}}))
=7\times 10^{-24}$, which agree to the simulation results. We see that the [[$5865,143,105$]] code
stands out because of its high code rate and extremely low logical error rate, making it a very promising code in practice.
\begin{figure}[!ht]
\centering\includegraphics[width=80mm]{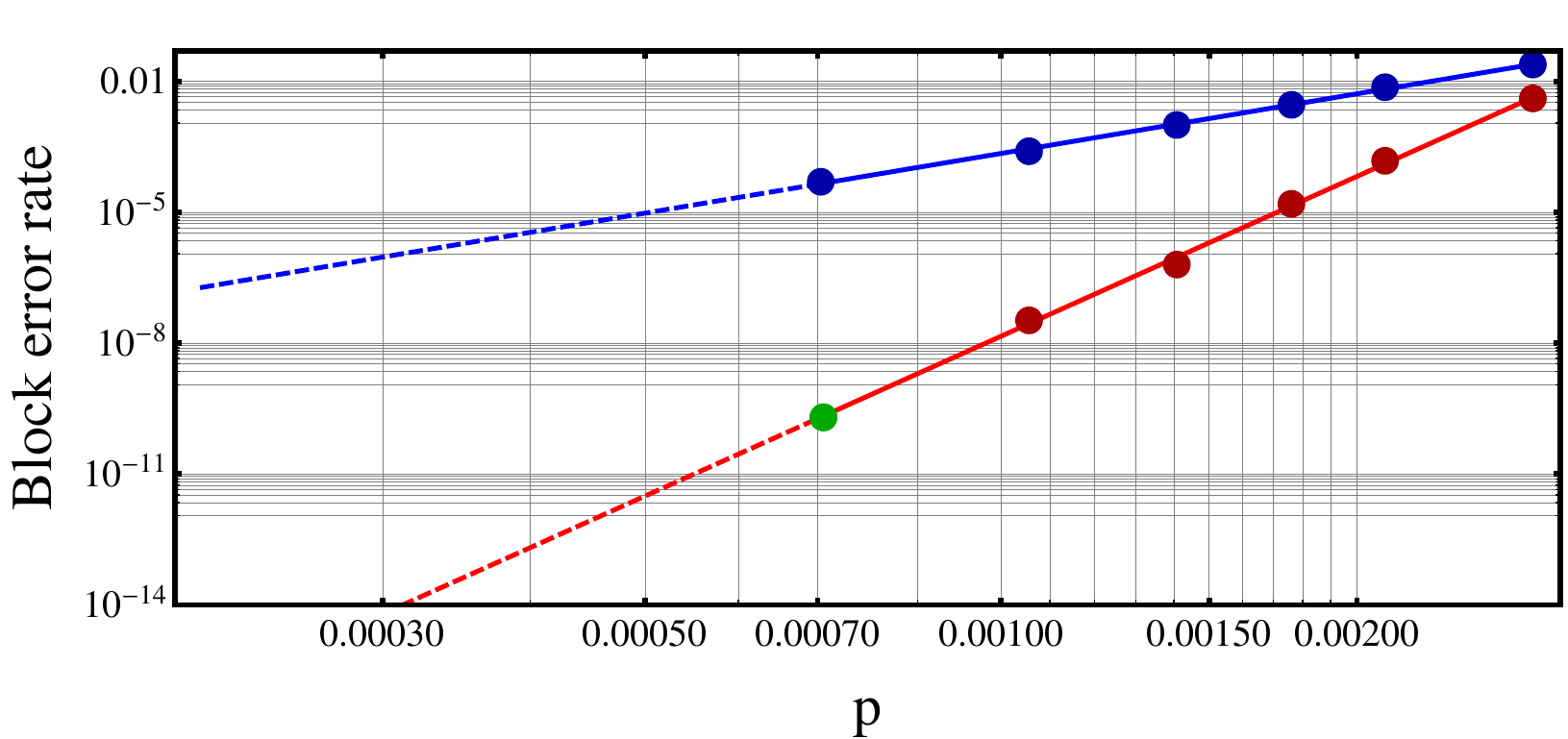}
\caption{\label{fig:processing_block} (Color online) The logical error rate of the logical $T$ gate performed on the concatenated [[$15,1,3$]] code of two levels (blue) and three levels (red)
using up to $3\times 10^7$ samples for each point. The green point represents a numerical upper bound for three levels when the physical error rate is $p=7\times 10^{-4}$.
The dashed lines are from extrapolation of linear fitting. }
\end{figure}
\vspace{-2mm}

For the processor block, we need a CSS code that allows a transversal non-Clifford gate such as the concatenated [[$15, 1, 3$]] shortened Reed-Muller code~\cite{knill1996threshold} or the 3D gauge color code~\cite{bombin2013gaugecolorcode}.
Here we simulate the concatenated [[$15, 1, 3$]] code, which allows a transversal $T$ gate.
This code has the ability to correct almost all bit-flip errors when $p_{\text{eff}}$ is small.
There exists an optimal efficient soft-decision decoder for concatenated codes~\cite{Poulin:2006:052333}, which diagnoses the error syndromes
using a message-passing algorithm~\cite{MacKay:2003:CambridgeUniversityPress}.
We performed Monte-Carlo simulations of the concatenated [[$15,1,3$]] code of two and three levels with the soft-decision decoder. The results are plotted in Fig.~\ref{fig:processing_block}.
Note that for three levels of concatenation, the logical error rate drops to less than $2\times 10^{-12}$ at $p_{\text{eff}}=0.007$ (by extrapolation).
Hence, when the physical error rate is less than $5\times 10^{-4}$, the logical error rate is below $2\times 10^{-12}$ for a single round of syndrome extraction and all logical gates.
In this case, the error rate for each logical operation is well below $10^{-10}$, which will allow interesting quantum algorithms that are impossible to run on classical computers. Nor do we have any reason to believe this is optimal: in all likelihood, better codes exist for both the storage and processor blocks.

\emph{Conclusion}.-- We have proposed a scheme for FTQC using large block codes as memory blocks and the concatenated [[$15,1,3$]] Reed-Muller code as processor block.
We showed that its logical error rate can be made low enough to implement practically interesting quantum algorithm with reasonable physical error rates.
The number of physical qubits required to protect a single logical qubit can be reduced from thousands of qubits to hundreds of qubits or less,
and no magic state distillation is needed.
It is very likely that good codes, such as quantum LDPC codes~\cite{camara2005constructionsLDPC,MacKay:2004:2315,tillich2014quantumLDPC}, may allow memory blocks
with even better performance~\cite{gottesman2013overhead}. The memory error rate could be further reduced if we exploited the correlations between effective errors before and after the syndrome extraction circuits.

On the other hand, the use frequency of each clean ancilla state varies dramatically. The ancillas $|0\>_L^{\otimes k}$ and $|+\>_L^{\otimes k}$ for syndrome extraction are used much more often than those for specific measurements, teleportation and logical gates. Thus, the distillation protocols should vary for different ancilla states to maximize the throughput of ancilla generation.
The total resources needed depends strongly on the details of the ancilla distillation protocols, which will be carefully investigated in our future work~\cite{Ching-yiAncillaDistillation}.

We thank Daniel Gottesman and  Jim Harrington for useful discussions, and the Oak Ridge National Lab for providing the Titan supercomputing resource. This work was supported in part by the IARPA QCS program; by HRL subcontract No. 1144-400707-DS; and by NSF Grant No. CCF-1421078.

%


\pagebreak

\widetext
\begin{center}
{\large \textbf{Supplementary Materials}}
\end{center}
\setcounter{equation}{0}
\setcounter{figure}{0}
\setcounter{table}{0}
\setcounter{section}{1}

\makeatletter
\renewcommand{\theequation}{S\arabic{equation}}
\renewcommand{\thefigure}{S\arabic{figure}}
\renewcommand{\bibnumfmt}[1]{[S#1]}
\renewcommand{\citenumfont}[1]{S#1}

\section{Phase gate and SWAP gate}
\begin{figure}[!ht]
\centering\includegraphics[width=190mm]{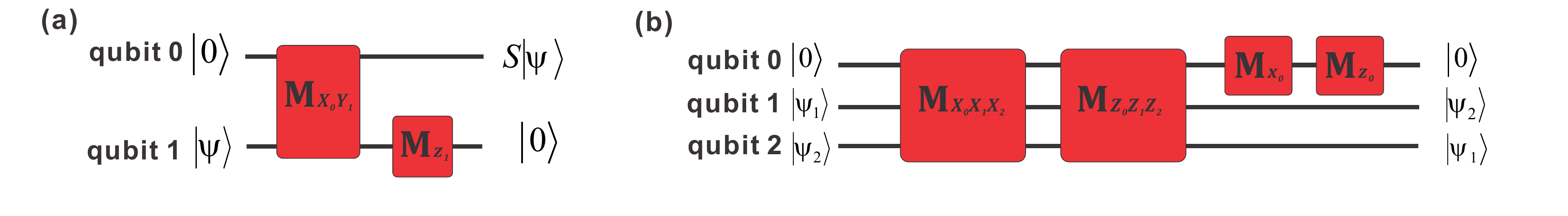}
\caption{\label{fig:phase_swap} (a) Phase gate. (b) SWAP gate. The red blocks represent joint measurements of logical qubits.}
\end{figure}

In this section, we show the details of implementing logical phase and SWAP gates in the same block using logical state measurements.
As shown in Fig.~\ref{fig:phase_swap}(a), assuming qubit 1 is in state $|\psi\>$, the phase gate can be realized by 1) preparing a buffer state $|0\>$ as qubit 0; 2) measuring $X_0Y_1$ followed by measuring $Z_1$. It will leave the state as $S|\psi\>|0\>$ up to a Pauli operator correction. For the SWAP gate, we prepare a buffer state initially in $|0\>$, as shown in Fig.~\ref{fig:phase_swap}(b). To SWAP $|\psi_1\>|\psi_2\>$, we can 1) measure $X_0X_1X_2$ followed by $Z_0Z_1Z_2$, followed by $X_0$. This will leave the state as $|+\>|\psi_2\>|\psi_1\>$ up to a Pauli operator correction. The buffer qubit can be reset to $|0\>$ by measuring $Z_0$.

\section{Effective error model}
There are three types of errors introduced by imperfect circuits in syndrome extraction and logical state measurement in our scheme. These are: measurement errors, gate errors, and preparation errors. Theorem~\ref{lemma:effective_error} states that it is possible to replace these errors with equivalent errors before and after the circuit if they are all \emph{independent} in the same block. We can then use these equivalent error processes as our error model and treat the circuit as being ideal. We treat these errors one at a time.

\begin{figure}[!ht]
\centering\includegraphics[width=90mm]{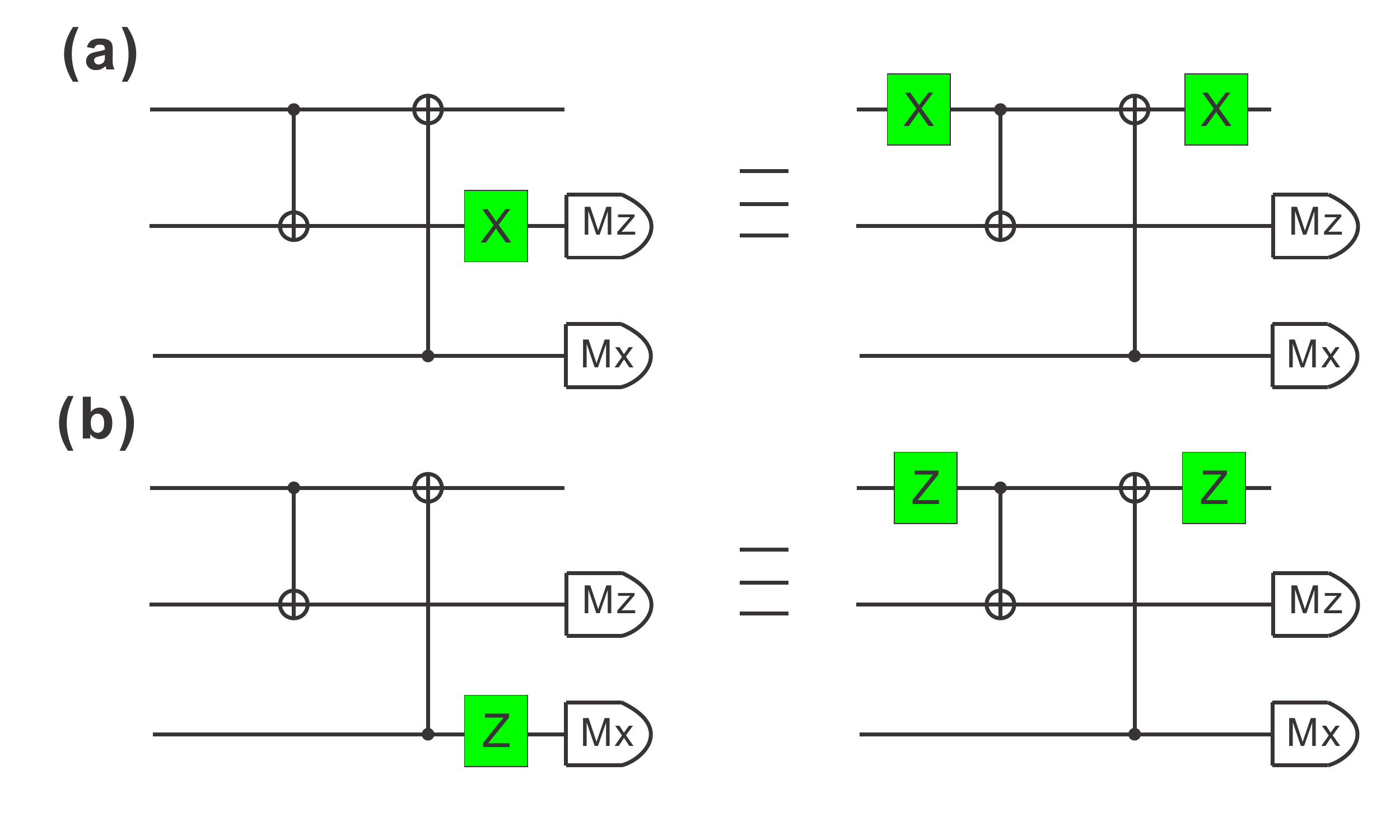}
\caption{\label{fig:measurement_error} The effect of measurement errors on the ancilla qubit can be effectively replaced by errors before and after the circuit.}
\end{figure}

Every error in a $Z$ measurement is equivalent to a single $X$ error followed by a perfect measurement, while every $X$ measurement error can be modeled by a single $Z$ error followed by a perfect measurement. In Fig.~\ref{fig:measurement_error},
we can see that an error in the $Z$ measurement is equivalent to two $X$ errors, before and after the circuit, on the codeword qubit. An error in an $X$ measurement has a similar effect.

\begin{figure}[!ht]
\centering\includegraphics[width=140mm]{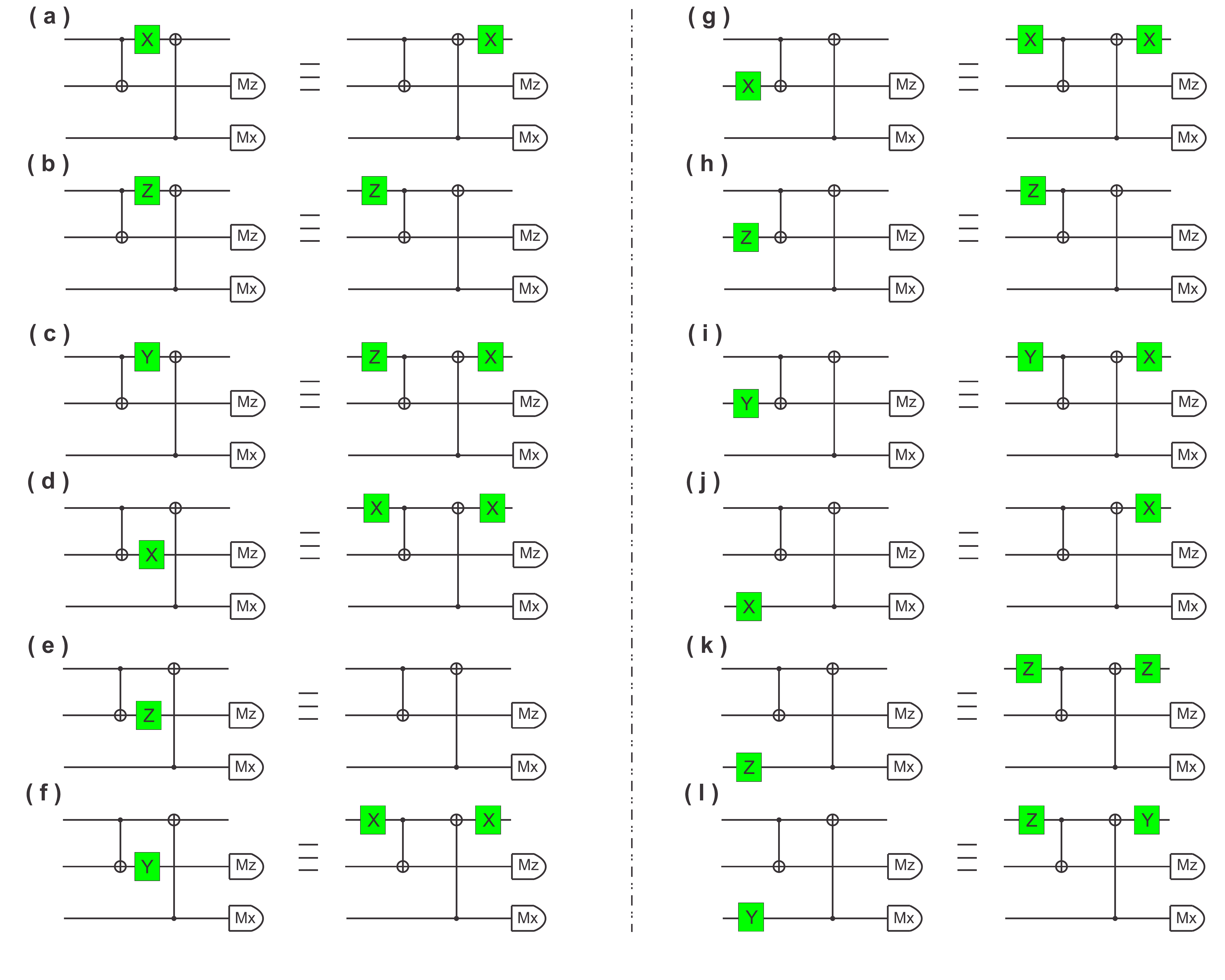}
\caption{\label{fig:gate_preparation error} The effect of gate errors (from (a) to (f)) on the first CNOT, and ancilla preparation errors (from (g) to (i)). Both types of errors can be replaced by errors before and after the circuit.}
\end{figure}

We treat a noisy gate as a perfect gate followed by an error. We treat the first CNOT explicitly here, and the situation for the second is just the same. An error in the CNOT gate is represented as a tensor product of two operators from $\{I,X,Y,Z\}$. Each such operator can be equivalently replaced by errors before and after the circuit. See panels (a) to (f) in Fig.~\ref{fig:gate_preparation error}.

Preparation errors include errors in initializing the ancilla blocks, and any memory or transport errors that occur in storing or distributing them. The replacement of preparation errors on each qubit of the two ancilla blocks is shown in panels (g) to (l) in Fig.~\ref{fig:gate_preparation error}.

\begin{figure}[!ht]
\centering\includegraphics[width=70mm]{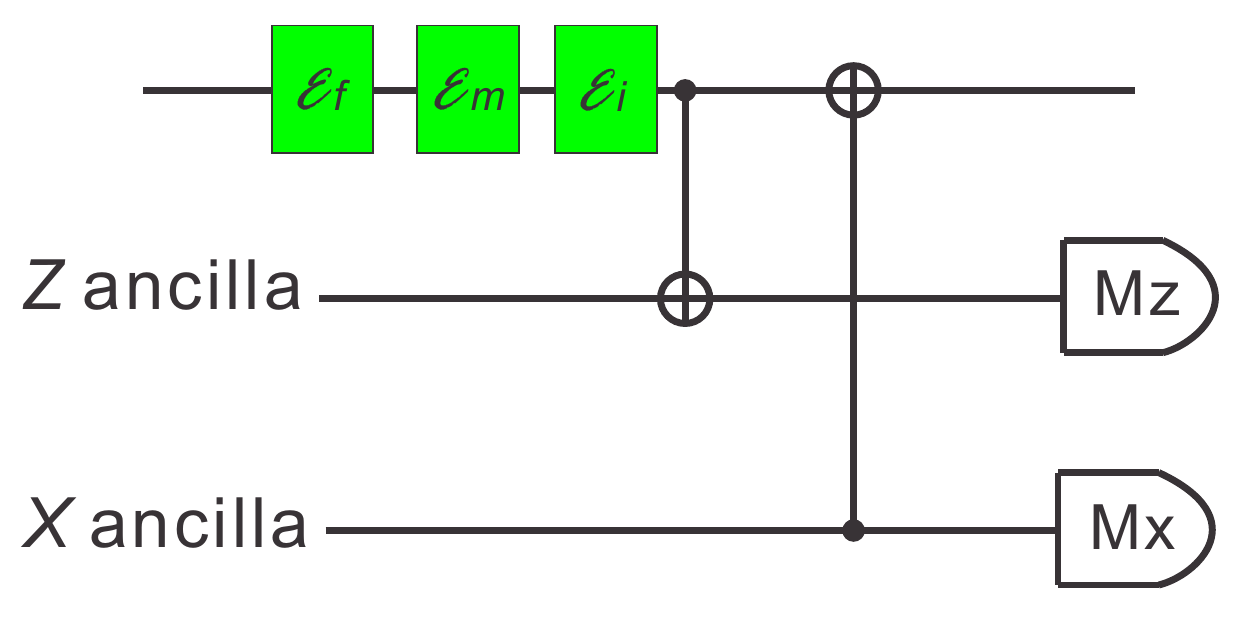}
\caption{\label{fig:effective_error_single_step}Effective error model in a single time step of circuit}
\end{figure}

From the argument above, we see that we can replace the noisy circuit with a perfect circuit preceded or followed by a noisy process on the codeword qubits at a single time step. Note that the errors before and after a circuit are \emph{correlated}. It may be possible to use this correlation to improve the estimation of the error, based on the entire time record of syndrome measurements. For now, we ignore this in finding the error process for a single time step. If there are also memory errors $\mathcal{E}_m$ on qubits in codeword block, then the error process before the circuit is
$\mathcal{E_\text{tot}}=\mathcal{E}_i\circ\mathcal{E}_m\circ \mathcal{E}_f$
as in Fig.~\ref{fig:effective_error_single_step}. Note that $\mathcal{E}_f$ here is the from previous circuit, and $\mathcal{E}_m$ is the memory error. For simplicity, suppose that all errors are Pauli errors, and define
$\mathcal{I}[\rho] = \rho,\  \mathcal{X}[\rho] = X\rho X, \ \mathcal{Z}[\rho] = Z\rho Z, \  \mathcal{Y}[\rho] = Y\rho Y$.
Since we model memory errors as depolarizing errors, $\mathcal{E}_m=(1-\epsilon)\mathcal{I}+\frac{1}{3}\epsilon(\mathcal{X}+\mathcal{Y}+\mathcal{Z})$.  $\mathcal{E}_f$ can be derived as follows:
\beq
\begin{split}
\mathcal{E}_f&=\underbrace{\left[(1-p_m)\mathcal{I}+p_m\mathcal{X}\right]\circ\left[(1-p_m)\mathcal{I}+p_m\mathcal{Z}\right]}_{\text{measurement}}\\
&\circ\underbrace{\left[\left(1-\frac{8}{15}p_{g_2}\right)\mathcal{I}+\frac{8}{15}p_{g_2}\mathcal{X}\right]}_{\text{first CNOT}}\circ\underbrace{\left[\left(1-\frac{12}{15}p_{g_2}\right)\mathcal{I}+\frac{4}{15}p_{g_2}(\mathcal{X}+\mathcal{Y}+\mathcal{Z})\right]}_{\text{second CNOT}}\\
&\circ\underbrace{\left[\left(1-\frac{2}{3}r\right)\mathcal{I}+\frac{2}{3}r\mathcal{X}\right]\circ\left[(1-r)\mathcal{I}+\frac{1}{3}r(\mathcal{X}+\mathcal{Y}+\mathcal{Z})\right]}_{\text{ancilla preparation}}.
\end{split}
\eeq
Similarly, we could have $\mathcal{E}_i$ as:
\beq
\begin{split}
\mathcal{E}_i&=\left[(1-p_m)\mathcal{I}+p_m\mathcal{X}\right]\circ\left[(1-p_m)\mathcal{I}+p_m\mathcal{Z}\right]\\
&\circ\left[\left(1-\frac{12}{15}p_{g_2}\right)\mathcal{I}+\frac{4}{15}p_{g_2}(\mathcal{X}+\mathcal{Y}+\mathcal{Z})\right]
\circ\left[\left(1-\frac{8}{15}p_{g_2}\right)\mathcal{I}+\frac{8}{15}p_{g_2}\mathcal{Z}\right]\\
&\circ\left[(1-r)\mathcal{I}+\frac{1}{3}r(\mathcal{X}+\mathcal{Y}+\mathcal{Z})\right]
\circ\left[(1-\frac{2}{3}r)\mathcal{I}+\frac{2}{3}r\mathcal{Z}\right].
\end{split}
\eeq
Thus, the entire noise process can be represented as:
\beq
\begin{split}
\mathcal{E}_{\text{tot}} &\equiv \mathcal{E}_i\circ \mathcal{E}_m\circ \mathcal{E}_f\\
&=\left(\frac{1}{3}\epsilon+\frac{4}{3}r+2p_m+\frac{16}{15}p_{g_2}\right)\mathcal{X}
+\left(\frac{1}{3}\epsilon+\frac{4}{3}r+2p_m+\frac{16}{15}p_{g_2}\right)\mathcal{Z}+
\left(\frac{1}{3}\epsilon+\frac{2}{3}r+\frac{8}{15}p_{g_2}\right)\mathcal{Y}\\
&+\left[1-\left(\epsilon + \frac{10}{3}r + 4p_m+\frac{8}{3}p_{g_2}\right)\right]\mathcal{I}+\mc{O}\left(\max\{\epsilon^2,r^2,p_m^2,p_{g_2}^2\}\right).
\end{split}
\eeq
If we set $\epsilon=p_m=r=p_{g_2}=p$, $\mc{E}_{\text{total}}$ can be approximated by $\mc{E}_{\text{total}}\approx (1-11p)\mc{I}+ \frac{71}{15}p\mathcal{X}+\frac{71}{15}p\mathcal{Z}+\frac{23}{15}p\mc{Y}$.

For non-Pauli errors, we can apply a similar trick by expanding the noise process in the Pauli basis and translating the errors term by term. In that case, each term also has a phase, which may depend on the measurement outcomes. This makes analysis more complicated but not really different in principle, especially since the syndrome measurements will tend to project into the Pauli basis.

\section{Stabilizer Formalism}\label{sec:stabilizer_formalism}
The $n-$qubit Pauli group $G_n$ is the set of all operators of the form
\beq
i^k O_1\otimes O_2\otimes\cdots \otimes O_n,
\eeq
where $O_j\in\left\{ I,X,Y,Z \right\}~\forall j$.

We ignore the overall phase $i^k$ in this section, which represents a global phase.
In general, for each $[[n,k,d]]$ stabilizer code with stabilizer group $S$, we can always do encoding operation from canonical code
$|\Psi\>=|0\>^{\otimes n-k}|\psi\>,$
where $|\psi\>$ is a $k-$qubit state is the quantum state to be protected and used in the computation. This is a stabilizer code, whose logical operators are $X_i$ and $Z_i$ for $i=n-k+1,\ldots, n$ and whose stabilizer generators are $Z_i$ for $i = 1,\ldots, n-k$. We can take the symplectic partners to be $X_i$ for $i = 1, \ldots, n-k$. This is the ``code" that we start with before the encoding operation. The encoding of $k$ qubits into $n$ qubits can be specified by a unitary operation $U_E\in\mathcal{C}_n$, where $\mathcal{C}_n$ is the normalizer of the Pauli group $G_n$ in $U(2^n)$ (Clifford group). Note that given a code, the encoding circuit is not unique. Here we choose a particular $U_E$. Under the encoding operation, the first $n-k$ $Z$ operators are mapped to $n-k$ stabilizer generators with relation $S_i = U_E Z_i U_E^\dag$, whereas the images of the $X$ operators acting on those qubits, $T_i= U_E X_i U_E^\dag$, are called pure errors or symplectic partners of $S_i$. The image of the Pauli operators acting on the last $k$ qubits are logical Pauli operators $\bar{X}_i=U_EX_{n-k+i}U_E^\dag$, $\bar{Z}_i=U_EZ_{n-k+i}U_E^\dag$.

Given a Pauli error $\mathscr{E}\in G_n$, we can always find the corresponding syndromes (assuming no measurement errors), so we define the syndrome extraction function $\mathcal{S}: G_n\rightarrow \{-1,1\}^{{n-k}}$,
where $\mathcal{S}(\mathscr{E})$ gives the syndromes of error $\mathscr{E}$. We can also define the following function: $\mathcal{T}: G_n\rightarrow G_n$, to represent pure error string. $\mathcal{T}(\mathscr{E})$ is uniquely determined by $\mathscr{E}$'s syndrome $\textbf{s}\equiv\mathcal{S}(\mathscr{E})=\{s_1,s_2\dots s_{n-k}\}$, and can be explicitly represented as:
\beq\label{eq:pure_error}
\mathcal{T}(\mathscr{E}) = U_{E}\left( \prod_{i=1}^{n-k} X_i^{-\frac{s_i-1}{2}}\right)U_{E}^{\dag},
\eeq
which is a product of pure errors. Define another function:
\beq
\mathcal{L}: G_n\rightarrow G_n, \ \ \mathscr{E}\mapsto \mathscr{E}\mathcal{T}(\mathscr{E}),
\eeq
where $\mathcal{L}(\mathscr{E})$ is a product of logical Pauli operators in the normalizer of stabilizer group $N(S)$. It is easy to observe that the $\mathscr{E}$ can be uniquely decomposed as:
\beq\label{eq:error_decomposition}
\mathscr{E} = \mathcal{L}(\mathscr{E})\mathcal{T}(\mathscr{E}).
\eeq
Last, we define a function which truncates the last $k$ single Pauli operators of a length $n$ Pauli operators string:
\beq
\mathcal{\mathscr{L}}:G_n\rightarrow G_k, \ \ \mathscr{E}\mapsto\mathscr{E}_{n-k+1}\cdots \mathscr{E}_{n}
\eeq
The images of the last $k$ $X$ and $Z$ Pauli operators of the canonical code correspond to the logical
operators of the code. If we run back the encoding circuit,  $\mathscr{L}\left(U_E^{\dag}\mathcal{L}(\mathscr{E})U_{E}\right)$ should indicate of logical errors actually happen according to $\mathscr{E}$. Note that if for $\mathscr{E}$ and $\mathscr{E}^\prime$, the corresponding $\mathcal{L}$ and $\mathcal{L}^\prime$ are different up to multiplication by some element in the stabilizer group, they should be regarded as equivalent. So when decoding, the only thing that matters is to estimate an equivalence class $\mathbb{L}$ containing all equivalent $\mathcal{L}$. Each equivalence class $\mathbb{L}$ corresponds to a unique $L\in G^k$, which is what logical error $\mathbb{L}$ stands for, and can be represented as $L=\mathscr{L}\left(U_E^{\dag}\mathcal{L}U_{E}\right)$ for any $\mathcal{L}\in \mathbb{L}$. So the optimal decoding problem can be transformed to a maximum \emph{a posterior} (MAP) probability problem:
\beq
\hat{L}=\arg\max_{L}P(L|\ \textbf{s}).
\eeq


For small quantum codes concatenated together (like the concatenated [[$15,1,3$]] code that is used in our scheme), an optimal and efficient decoder called the soft decision decoder does exist~\cite{Poulinsoft:2006:052333}. Consider an $l-$level concatenation code, and let the $j$th level be an [[$n_j,k_j,d_j$]] code with $k_j = 1$ for $j\neq 1$. The parameters of the corresponding concatenated code are [[$\prod_{j=1}^ln_j, k_1, \prod_{j=1}^ld_j$]]. Define $s_j^{(i)}\in\{-1,1\}^{n_j-k_j}$ be the syndromes of $i$th block of the $j$th concatenation layer after syndrome extraction . Denote $\textbf{s}_j^{(i)}$ be the collection of syndromes whose stabilizer generators act nontrivially on all \emph{physical} qubits associated to the $i$th block of the $j$th concatenation layer. In other words, these sets of syndromes can be defined as: $\textbf{s}_j^{(i)}=\left\{s_j^{(i)}\right\}\mathlarger{\bigcup} \left\{\mathlarger{\bigcup}_{p=in_j-i+1}^{in_j}\textbf{s}_{j+1}^{(p)}\right\}$. Note that $\textbf{s}_l^{(i)}=s_l^{(i)}$ are the syndromes of codes in bottom level.  At last, denote $\textbf{s}_j=\mathlarger{\bigcup_{i=1}^{\prod_{q=1}^{j-1}n_q}}\textbf{s}_{j}^{(i)}$ to be the collection of all the syndromes from the layers $j$ to $l$.
Then it is easy to see that $\textbf{s}_1$ is the set of all syndromes according to the concatenated code's stabilizer generators. Estimating $L$ for the concatenated code is equivalent to estimating $L_1$ for the code at the top level. The decoding is equivalent to finding $\arg\max_{L_1}P(L_1|\textbf{s}_1)$. Define $\textbf{L}_2=(L_2^{(1)}\ldots L_2^{(n_1)})$ to be an array of logical operators at the second level. Then this probability can be factorized by conditioning on $\textbf{L}_2$,
\beq\label{eq:soft_decoder}
\begin{split}
P(L_1|\ \textbf{s}_1)
&=\mathlarger{\sum}_{\textbf{L}_2}P(L_1|\ \textbf{s}_1, \textbf{L}_2)P(L_2|\ \textbf{s}_1)\\
&=\mathlarger{\sum}_{\textbf{L}_2}\frac{\delta[L_1=\mathscr{L}_1\big(U_E^{1\dag}
\textbf{L}_2 U^1_{E}\big)]\ \delta[s_1=\mathcal{S}_1(\textbf{L}_2)]}{P(s_1|\ \textbf{s}_2)}\ \mathlarger{\prod}_{i=1}^n P(L_2^{(i)}|\ \textbf{s}_2^{(i)}),
\end{split}
\eeq
where $L_j$, $\mathscr{L}_j$, $U_E^j$ and $\mathcal{S}_j$  are $L$, $\mathscr{L}$, $U_E$ and $\mathcal{S}$ corresponding to the code at the $j$th level, and $\delta(\cdot)$ is the indicator function. The derivation repeatedly uses the Bayes rule and the following facts:
a) The syndromes and logical errors of level $j$ are determined by the logical errors of layer $j+1$. b) The channel is not correlated (or more specifically, that the error model can only correlate qubits in the same block but not across different blocks).
The optimal decoding is reduced to a {\rm sum-product} problem on a tree which can be exactly and efficiently solved using message passing algorithm. If the estimated logical error is $\hat{L}_1=\mathscr{L}(U_E^\dag\mathcal{L}(\mathscr{E})U_E)$, the decoding is a success, otherwise it is a failure. The error at the physical level then can be estimated as:
\beq\label{eq:soft_decoder}
\hat{\mathscr{E}}=U_E\ \left(I^{\otimes n-k}\otimes\hat{L}_1  \right)\ U_E^{\dag}\mathcal{T}(\mathscr{E}),
\eeq
which is used for actual correction at the physical level.

In general, the optimal decoding problem is NP hard~\cite{Berlekamp:1978:384}. In practice, one must use codes with some structure (like the ones chosen for the memory block) so that an efficient hard decision decoding algorithm provides an error estimate $\hat{\mathscr{E}}$ based on the assumption that errors are all independent and the same type of error is equally likely to happen to different qubits. Efficient hard decision decoders for quantum BCH codes and the quantum Golay exist. The Golay code can be decoded up to its correctability $ t= \lfloor \frac{d-1}{2}\rfloor = 3$ using the Kasami error-trapping decoder. The BCH codes can be decoded using the Berlekamp-Massey algorithm . We can decompose $\hat{\mathscr{E}}$ into $\hat{\mathscr{E}}=\mathcal{L}(\hat{\mathscr{E}})\mathcal{T}_{\textbf{s}}$,
where $\mathcal{T}_{\textbf{s}}=\mathcal{T}(\mathscr{E})=\mathcal{T}(\hat{\mathscr{E}})$. If $\hat{L}=\mathscr{L}\left(U_E^\dag\mathcal{L}(\hat{\mathscr{E}})U_E\right)$ is equal to $L=\mathscr{L}\left(U_E^\dag\mathcal{L}(\mathscr{E})U_E\right)$, we declare that the error correction is correct, otherwise, we declare that a failure.

To determine whether a correction of a random Pauli error is correct, we need to run the encoding circuit backward. So first of all, we need to find the encoding circuit of a stabilizer code. Using the symplectic form, this can be done in the following way.
Let the symplectic matrix representation of the $n-k$ stabilizer generators and logical operators be ${\bf M}$.  A Clifford circuit is an automorphism of Pauli group $G_n$, which can be represented as a linear map that acts on the matrix:
\beq
{\bf M}\rightarrow {\bf M}^\prime={\bf MC},
\eeq
where ${\bf C}$ is a nonsingular $2n\times 2n$ binary matrix representing the action of the Clifford circuit. Since ${\bf C}$ represents a unitary transformation, it must preserve the commutation relations of the operators it transforms; this restriction corresponds to the constraint
\beq
{\bf CJC}^T={\bf J} \ \ \text{with} \ \ {\bf J}=\left[
                                           \begin{array}{cc}
                                             {\bf 0} & {\bf I} \\
                                             {\bf I} & {\bf 0} \\
                                           \end{array}
                                         \right].
\eeq
Note that the $n+k$ rows of the matrix {\bf M} do not form a full basis. This reflects the fact that $n+k$ Pauli operators they represent do not form a full set of generators. We can supplement them by adding an additional $n-k$ symplectic partners of the stabilizer generators  logical operators to {\bf M} to form a new full rank matrix {\bf M}.

The symplectic partner $T_i$ of $S_i$ should satisfy
\beq
[T_i, S_j] = 0 \ \ \text{for } \ \ i\neq j, \ \ \ \{T_i, S_i\} = 0.
\eeq
Start from the canonical code we described before, and define the matrix ${\bf M}_0$ to represent this ``code". We use the following order for the operators: $X_n,\ldots, X_1,$ $Z_n,\ldots Z_1$. This gives us the very simple matrix
\beq
{\bf M}_0=\left[
            \begin{array}{cc|cc}
              {\bf I}_{n-k} & {\bf 0} & {\bf 0} & {\bf 0} \\
              {\bf 0} &  {\bf I}_{k} & {\bf 0} & {\bf 0} \\
              {\bf 0} & {\bf 0} & {\bf I}_{n-k} & {\bf 0} \\
              {\bf 0} & {\bf 0} & {\bf 0} &  {\bf I}_{k} \\
            \end{array}
          \right]\equiv {\bf I}_{2n}
\eeq
where the first $n-k$ rows are the symplectic partners, the next $k$ rows are the logical $X$ operators, the next $n-k$ rows are the stabilizer generators and the last $k$ rows are the logical $Z$ operators.

From this code, we produce the $[[n,k,d]]$ code we are interested in by applying an encoding circuit $U_E$, which has representation ${\bf C}_E$:
\beq
{\bf M} = {\bf M}_0 {\bf C}_E ={\bf C}_{E}.
\eeq
In other words, if we know all logical operators and symplectic partners of the stabilizer code, we can build the encoding matrix of that code directly. Suppose we know all of the stabilizer generators and logical $Z$ operators and $\bar{X}_l$, for $l = 1,\cdots i$, we can recursively find $\bar{X}_{i+1}$ by solving the following linear equation :
\beq
\begin{split}
&\left[
  \begin{array}{ccc|ccc|ccc}
     \bar{X}_i^\text{t}& \cdots & \bar{X}_1^\text{t} & S_{n+k}^\text{t} & \cdots & S_1^\text{t} & \bar{Z}_k^\text{T} & \cdots & \bar{Z}_1^\text{T} \\
  \end{array}
\right]^\text{t}\odot \bar{X}_{i+1}\\
&= \left[\ \underbrace{0 \ \ \cdots \ \ 0}_{i} \ | \underbrace{\ 0 \ \ \cdots \ \ 0}_{n+k} \ | \underbrace{\ 0 \ \ \cdots \ \ 0}_{k-i-1} \ 1 \ \underbrace{0 \ \ \cdots \ \ 0}_{i} \
\right]^\text{t},
\end{split}
\eeq
where $\odot$ is the symplectic inner product between binary strings. Similarly, if we find all $S_i$, $\bar{Z}_i$, $\bar{X}_{i}$ and $T_l$ for $l=1\cdots i$, we could find $T_{i+1}$ by solving the following equation:
\beq
\begin{split}
&\left[
  \begin{array}{ccc|ccc|ccc|ccc}
 T^\text{t}_i & \cdots & T^\text{t}_1 &\bar{X}_k^\text{t}& \cdots & \bar{X}_1^\text{t} & S_{n+k}^\text{t} & \cdots & S_1^\text{t} & \bar{Z}_k^\text{t} & \cdots & \bar{Z}_1^\text{t} \\
  \end{array}
\right]^\text{t}\odot T_{i+1}\\
&= \left[\ \underbrace{0 \ \ \cdots \ \ 0}_{i} \ | \underbrace{\ 0 \ \ \cdots \ \ 0}_{k} \ | \underbrace{\ 0 \ \ \cdots \ \ 0}_{n+k-i-1} \ 1 \ \underbrace{0 \ \ \cdots \ \ 0}_{i} \ | \ \underbrace{0 \ \ \cdots \ \ 0}_{k} \
\right]^\text{t}.
\end{split}
\eeq

\section{Quantum BCH code and Quantum Golay code}
Since the size of a concatenated code is the product of the code sizes at each layer, choosing codes that are not too big is important if one wishes to concatenate them. The Golay code and the BCH codes~\cite{shu2004error} are well-known classical cyclic codes due to their remarkable algebraic structures and the ability to decode multiple errors that is especially useful when the size of the code is small.
The scheme we consider for protecting the memory blocks against error is to use two-layer concatenation of a quantum BCH code at the top layer and the [[$23,1,7$]] quantum Golay code at the bottom layer. The quantum codes that we consider here are derived from their classical counterparts by the CSS construction. We pick the following three quantum BCH codes constructed from self-dual classical BCH codes that have reasonable code lengths, good code rates and good code distances:

1) [[$89,23,9$]] quantum BCH code, which is derived from the [$89,56,9$] classical BCH code. The generator polynomial of the classical code is $X^{33}+X^{30}+X^{27}+X^{26}+X^{25}+X^{24}+X^{22}+X^{21}+X^{20}+X^{16}+X^{15}+X^{14}+X^{11}+X^{10}+X^9+X^6+X^3+X^2+1$.

2) [[$127,57,11$]] quantum BCH code, which is derived from the [$127,92,11$] classical BCH code. The generator polynomial of the classical code is $X^{35}+X^{34}+X^{33}+X^{28}+X^{24}+X^{23}+X^{22}+X^{19}+X^{17}+X^{15}+X^{12}+X^{11}+X^9+X^8+X^6+X^4+X^2+X+1$.

3) [[$255,143,15$]] quantum BCH code, which is derived from the [$255,199,15$] classical BCH code. The generator polynomial of the classical code is $X^{56}+X^{51}+X^{50}+X^{49}+X^{46}+X^{43}+X^{41}+X^{40}+X^{39}+X^{34}+X^{30}+X^{26}+X^{25}+X^{24}+X^{22}+X^{20}+X^{17}+X^{16}+X^{11}+X^{10}+X^8+X^7+X^4+X^3+X^2+X+1$.

Note that the [[$23,1,7$]] quantum Golay code is derived from the [23,12,7] classical Golay code. The generator polynomial of the classical code is $X^{11}+X^{10}+X^6+X^5+X^4+X^2+1$.

In all four cases, logical $Z_i$ operators can be obtained by shifting the corresponding generator polynomials for classical codes.

\section{[[$15,1,3$]] Reed-Muller Code}
The concatenated [[$15,1,3$]] code is constructed from the truncated classical Reed-Muller code.
Classical Reed-Muller codes are weakly self-dual codes
with simple and good structural properties \cite{MacWilliams:1983:NorthHolland}.
A Reed-Muller code has two
parameters $r,m$, $0\leq r\leq m$,
and is denoted by $RM(r,m)$.
This code is of length $2^m$ and $r$ is called its order. Let $C=RM(1,4)$ with parameters $[16,5,8]$.
Consider the following $(m+1)$ $2^m$-tuples:
{ \[
\begin{array}{cccccccccccccccc}
v_0=\mathbf{1} &=&[&1111 & \cdots & 1111 & 1111 & \cdots & 1111&],\\
       v_1 &=& [&0101 & \cdots & 0101 & 0101 & \cdots & 0101&],\\
       v_2 &=& [&0011 & \cdots & 0011 & 0011 & \cdots & 0011&],\\
      &\vdots&  &   \vdots  & \ddots & \vdots &     \vdots & \ddots & \vdots &\\
       v_m &=& [&0000 & \cdots & 0000 & 1111 & \cdots & 1111&],
\end{array}
\]}
$C$ is generated by $v_0$, $v_1$, $v_2$, $v_3$ and $v_4$. The codewords of $C$ have weight divisible by 8. Let $C^\prime$ be the code of $C$ shortened by deleting the first bit.
Then its codewords have weight either 0 or 7 mod 8. Let $v_j^\prime$ be the punctured generator by deleting the first bit of $v_j$, and then $C^\prime$ is a $[15,5,7]$ code generated by $v_0^\prime$, $v_1^\prime$, $v_2^\prime$, $v_3^\prime$ and $v_4^\prime$.
Let $C_0^\prime$ be the even subcode of $C^\prime$ generated by $v_1^\prime$, $v_2^\prime$, $v_3^\prime$, $v_4^\prime$. Note that $v_1$, $v_2$, $v_3$ and $v_4$ have leading bit 0, but $v_0$ has a leading bit 1. $C_0^\prime$ has parameters $[15,4,8]$ and its codewords have weight divisible by 8. The $[[15,1,3]]$ quantum code is encoded as following:
\beq
|0\>_L =\mathlarger{\sum}_{u\in C_0^\prime}|u\>,\ \ \ \text{and} \ \ \ |1\>_L =\mathlarger{\sum}_{u\in C_0^\prime}|u+v_0\>.
\eeq
The corresponding parity check matrix for the $[[15,1,3]]$ code is
\[
\begin{bmatrix}
000000011111111& \\
000111100001111& \\
011001100110011& \\
101010101010101& \\
&000000011111111 \\
&000111100001111 \\
&011001100110011 \\
&101010101010101 \\
&000000000001111 \\
&000000000110011 \\
&000000001010101 \\
&000001100000011 \\
&000010100000101 \\
&001000100010001 \\
\end{bmatrix},
\]
which is asymmetric between $X$ and $Z$.
Then
\beq
\begin{split}
T^{\dag \otimes n}|x\>_L=& \sum_{u\in C_0'} T^{\dag \otimes n} |u+ x v_0\> = \sum_{u\in C_0'} e^{-i\frac{ \wt{u+xv_0} \pi}{4}} |u+ x v_0\>\\
=& e^{-i\frac{ 7x \pi}{4}} \sum_{u\in C_0'}  |u+ x v_0\> = e^{-i\frac{ 7x \pi}{4}} |x\>_L,
\end{split}
\eeq
which implements the logical $T$ gate. The third equality follows because for any $v\in C_0^\prime$, ${\rm wt}(v)=0$ mod 8, while for any $v\in C^\prime_0+v_0$, ${\rm wt}(v)=7$ mod 8.

This code has a particularly large ability to correct almost all bit-flip errors when $p_{\text{eff}}$ is small.
Monte-Carlo simulation of two and three level concatenation of [[$15,1,3$]] using soft decision decoder has been shown in Fig.~\ref{fig:processing_block}. The simulation of 3-level concatenation using the soft-decision decoder is quite time consuming. It was implemented on the Titan supercomputer at Oak Ridge National Lab. We use $10^6$ samples for  $p_{\text{eff}}=0.04$, $10^7$ samples for
$p_{\text{eff}}=0.03$,  $3\times10^7$ samples for
$p_{\text{eff}}=0.02$ and  $3\times10^7$ sample for
$p_{\text{eff}}=0.015$.

Note that for $p_{\text{eff}}\leq 0.01$, the logical error rate is so small that the number of samples needed is too large for simulation. Even on Titan, direct Monte Carlo simulation is much too expensive. We need to be careful here, since we are using a message passing algorithm, so the error floor effect in classical message passing decoding might potentially occur in the low error rate region. To evaluate the performance in the region $p_{\text{eff}}\leq 0.01$, we can use the following observation. The distance for $X$ errors in the 3-level concatenated [[$15,1,3$]] code ([[$3375,1,27$]]) is 343, which means the code can correct all $X$ error of weight less than 170 since the decoder is optimal. So we just need to focus on $Z$ errors. Define $P_L(w,p_{\text{eff}})$ as the logical error probability after soft decision decoding using parameter $p_{\text{eff}}$ when Pauli errors of weight $w$ uniformly occur on each qubit. Define $P^Z_L(w,p_{\text{eff}})$ as the logical error probability when $Z$ errors of  weight $w$ uniformly occur on each qubit. Consider an all-$Z$ error $E$ of weight $w<170$ on a set of qubits $\mathscr{S}$. Let $E^\prime$ be arbitrary string of $X$ errors supported by $\mathscr{S}$. If we can correct $E$, then $E^\prime\cdot E$ can be corrected. Thus, we have $P_L(w,p_{\text{eff}})=\left(\frac{2}{3}\right)^wP^Z_L(w,p_{\text{eff}})$ for $w<170$. Then the logical error probability using soft-decision decoder can be bounded as follows:
{\small
\beq
\begin{split}
P_L(p_{\text{eff}})&=\mathlarger{\sum}_{w=1}^{3375}P_L(w,p_{\text{eff}})\binom{3375}{w}p_{\text{eff}}^w(1-p_{\text{eff}})^{3375-w}\\
                   &=\mathlarger{\sum}_{w=14}^{34}\binom{3375}{w}\left(\frac{2}{3}\right)^wP^Z_L(w,p_{\text{eff}})p_{\text{eff}}^w(1-p_{\text{eff}})^{3375-w}
                   +\mathlarger{\sum}_{w=35}^{50}\binom{3375}{w}\left(\frac{2}{3}\right)^wP^Z_L(w,p_{\text{eff}})p_{\text{eff}}^w(1-p_{\text{eff}})^{3375-w}
                   \\
                   &+\mathlarger{\sum}_{w=51}^{100}\binom{3375}{w}\left(\frac{2}{3}\right)^wP^Z_L(w,p_{\text{eff}})p_{\text{eff}}^w(1-p_{\text{eff}})^{3375-w}
                   +\mathlarger{\sum}_{w>101}\binom{3375}{w}P_L(w,p_{\text{eff}})p_{\text{eff}}^w(1-p_{\text{eff}})^{3375-w}\\
                   &\leq\mathlarger{\sum}_{w=14}^{34}\left(\frac{2}{3}\right)^w\binom{3375}{w}P^Z_L(34,p_{\text{eff}})p_{\text{eff}}^w(1-p_{\text{eff}})^{3375-w}
                   +\mathlarger{\sum}_{w=35}^{50}\left(\frac{2}{3}\right)^w\binom{3375}{w}P^Z_L(50,p_{\text{eff}})p_{\text{eff}}^w(1-p_{\text{eff}})^{3375-w}
                   \\
                   &+\mathlarger{\sum}_{w=51}^{100}\left(\frac{2}{3}\right)^w\binom{3375}{w}p_{\text{eff}}^w(1-p_{\text{eff}})^{3375-w}
                   +\mathlarger{\sum}_{w>100}\binom{3375}{w}p_{\text{eff}}^w(1-p_{\text{eff}})^{3375-w}
\end{split}
\eeq
}
We evaluate $P_L^Z(34,p_{\text{eff}})$ and $P_L^Z(50,p_{\text{eff}})$ for $p_{\text{eff}}=0.01$ and get an upper bound on the logical error probability of $2\times 10^{-10}$.

%

\end{document}